\documentstyle[pra,aps,multicol,psfig]{revtex}

\def\addcontentsline#1#2#3{\relax}
\begin{document}
\draft
\title{Transport through a 1D Mott-Hubbard insulator of finite length. }
\author{V.V. Ponomarenko$^{*}$ and N. Nagaosa }
\address{Department of Applied Physics, University of Tokyo,
Bunkyo-ku, Tokyo 113, Japan}
\date{\today}
\maketitle
\begin{abstract}
Transport through a 1D Mott-Hubbard insulator of a finite length $L$
is studied beyond perturbative approach.  At special value of the 
low energy constant of the interaction we have mapped the problem onto 
the exactly solvable models and found current vs. voltage $V$ at high
temperature $T>max(m,T_L)$ and at low energy $T,V<T_L$ 
($T_L=v_c/L$; $v_c:$ charge velocity). The result shows 
that for the strong interaction creating a large Mott-Hubbard gap 
$2m \gg T_L $   inside the wire,
the transport is suppressed  near half-filling 
everywhere inside the gap except for an exponentially small 
region of $V,T < T_L exp(-2m/T_L)$.

\end{abstract}


%
\multicols{2}
Motivated by current experiments\cite{Tar} of Tarucha's group, 
a few works recently appeared where
an effect of Umklapp processes on quantum transport
through a 1D wire has been studied. 
Previously, it has been already shown by experiment \cite{tar} 
and in theory\cite{mas1,pon,safi} that 
there occurs no renormalization of the conductance due to 
the electron-electron interaction. Nonetheless, this 
renormalization has been observed \cite{ya}. 
A recent calculation\cite{kami} based
on Kubo formalism \cite{kawa} studied the effect of the Umklapp 
backscattering and concluded that it affects the zero temperature
conductance. This contradicts the
perturbative results of Ref. \cite{us,od}, where the quantized
conductance $G= e^2/(\pi \hbar)$ has been obtained at $T=0$ 
(below we use the units, where $e=\hbar =1$).  
Another point of view suggests \cite{fuk,mas3} that
in a non-perturbative case of large Mott-Hubbard gap 
$2m \gg v_c/L \equiv T_L$
($L$: length of the wire; $v_c$: charge velocity) the conductance
is strongly suppressed at low energy inside the gap similar to the band gap 
case. 

In this paper, to clear up the difference between the Mott-Hubbard insulator 
and the band
gap one, we map the problem at low energies and at high 
temperature onto the exactly solvable models making use of a free fermion
value of the constant $g$ of the forward scattering inside the wire. 
The results are shown in Figs. 1 and 2. At low energies when 
$T,V \ll T_L$ 
($T$: temperature; $V$: voltage), we have found that  
a new energy scale $T_x \propto T_L exp[-2\sqrt{m^2-\mu^2}/T_L]$ 
appears in the system 
if the chemical potential $\mu$ is small enough:  
$\sqrt{m^2-\mu^2}/T_L \gg 1$. 
Below $T_x$ the conductance is not suppressed contrary to the expectation of 
Starykh and Maslov \cite{mas3}
and the current increases linearly.
Above this energy the current saturates and the conductance goes down as $T_x/T$
reaching small values $\approx exp[-2\sqrt{m^2-\mu^2}/T_L]$ at $T \approx T_L$. 
At high temperature
$T \gg T_L,m$ we confirmed the asymptotical behavior of conductance: 
$G=(1-cst {m \over T} (1+\cosh{\mu \over T})^{-1} )/\pi$ predicted before for 
Mott-Hubbard insulator in perturbative regime of small gap \cite{us,od}
where the constant depends on $m/T_L$.

A brief physical explanation to these results follows. 
At low energies $T < T_L$ and $\mu \ll m$ the charge field is quantized
inside the wire at its values related to the degenerate
sin-Gordon vacua. Rare low energy excitations tunnel through the wire
with the amplitude $\propto exp[-m/T_L]$ as (anti)solitons
switching the quantized value of the field. 
The whole process of tunneling, however, includes
transformation of the reservoir electron into the sin-Gordon quasiparticles
and back. This transformation results in a non-trivial scaling dimension
of the tunneling operator equal to  $1/2$ for the Mott-Hubbard insulator 
connected to the Fermi liquid reservoirs
independently of any parameters. 
In the case of the band insulator, this dimension is marginal ($=1$): 
the transformation is trivial and does not introduce additional
energy dependence.
The infrared relevantness of the tunneling with the $1/2$ dimension brings
out above resonance at zero energy. Meanwhile, the exponentially small tunneling
amplitude specifies the narrow width of  this resonance equal to
the crossover energy. Increase of $|\mu |$ favours tunneling of the
quasiparticles of the same sort 
and ultimately produces their finite density in the wire.
Then the interaction between these quasiparticles described with
the two-particle $S$-matrices \cite{zam} dependent on $g$ emerges. At low momenta
the $S$-matrix for the quasiparticles of the same sort is inevitably free fermion
like, as at $g=1/2$. 
It manifests in the renormalization group (RG) flow derived from the 
Bethe anzats solution for the massive phase \cite{ktrg}  of the sin-Gordon model
and in the exponent calculated for the Tomonaga Luttinger liquid
(TLL) phase at low density \cite{onehalf}. Increase of $T$, on the other
hand, is expected to entail, first, a thermally activated behavior of the
conductance $\propto exp[-2m/T]$ at $T_L<T<m$ \cite{rice} and then a
power law dependence at $m<T$. Since the effective value of $g$, in general,
scales with energy,
the $1/T$ dependence we found for
$g=1/2$ may vary at higher energies $T \gg m$ depending on the high energy value 
of $g$.

Transport through the finite length wire under a constant voltage $V$ between
the left and right leads could be described in the inhomogeneous
Tomonaga-Luttinger liquid model (TLL) with the Lagrangian \cite{us,us2}:
$\int dx \{ \sum_b {\cal L}_b(x,\phi_b,\partial_t \phi_b) + 
{\cal L}_{bs}(x,Vt,\phi_c, \phi_s) \}$. The 
bosonic fields $\phi_b(x,t), b=c,s$  relate to the deviations of 
the charge and
spin densities from their average values as following: 
$\rho_{b}(x,t)=(\partial_x \phi_{b}(x,t))/(\sqrt{2} \pi)$,
respectively. The first part of the Lagrangian describes a
free electron motion modified by the forward scattering interaction.
The second part of the Lagrangian
introduces backscattering inside the wire.
Only its term corresponding to the Umklapp process of four
Fermi momenta transfer 
is important
near half-filling. This term does not involve
the spin field. Therefore, our consideration will be restricted to the charge 
field only. For the clean wire this field is characterized by the Lagrangian:
\endmulticols
\vspace{-6mm}\noindent\underline{\hspace{87mm}}
\begin{equation}
\int d x {\cal L}_t= \int dx \bigl[ \frac{v_c(x)}{2g(x)}
\{ 
{1 \over v_c^2} \left({{\partial_t \phi_c(t,x) }
     \over {\sqrt{4 \pi}}}
\right)^2 -
\left({{\partial_x \phi_c(t,x) } \over {\sqrt{4 \pi}}} \right)^2 
\}
-
{E_F^2 U \over v_F} \varphi(x) \cos(4 k_{F} x + 2 V t +
\sqrt{2} \phi_c(t,x)) \bigr]
\label{1}
\end{equation}
\noindent\hspace{92mm}\underline{\hspace{87mm}}\vspace{-3mm}
\multicols{2}
\noindent
where $\varphi(x)= \theta (x) \theta (L-x)$ specifies a one channel wire
of the length $L$ adiabatically attached to the leads $x>L,x<0$
and $v_F(E_F)$ denotes the Fermi velocity(energy) in the channel. 
The parameter $4k_F$ varies the chemical potential $\mu=2k_F v_c=E_{thr}$
of the wire from its zero value at half-filling. Outside the
Hubbard gap this parameter coincides with the momentum
transfered by the backscattering: four Fermi momenta minus a vector of 
the reciprocal lattice, and relates the present results to the 
perturbative ones \cite{us}.
The constant of the forward scattering varies from 
$g_c(x)=g$  inside the wire  ($x \in [0,L]$) to 
$g_c(x) \equiv g_\infty =1$  inside the leads, and the Umklapp scattering of 
the strength $U$ is introduced inside the wire. The charge velocity
$v_c(x)$ changes from $v_F$ outside the wire to a some constant $v_c$
inside it. In the absence of the Umklapp scattering, $v_c \simeq v_F/g$ and $0<g<1$
is determined by the forward scattering amplitude of the bare short range 
interaction between electrons. Approaching the half-filling put the
Umklapp scattering on. It entails an essential renormalization of the 
low energy value of $g$, which flows to its free fermion value $g=1/2$ 
in the massive phase \cite{ktrg} $(\mu < m)$ 
where the coefficient of the $\cos $-term scales to $\simeq m^2$
and on approaching 
this phase \cite{onehalf} $|\mu | \searrow m$. 
This value of $g$ will be assumed below.
The zero frequency current through 
the wire equals $I=V/\pi + <\hat{I}_{bsc}>$, where the backscattering current 
\cite{us2} is
$\hat{I}_{bsc}=-2 E_F^2 U/v_F \int_0^L dx \sin(4k_F x +2Vt +\sqrt{2} \phi_c(x))$.
It will be shown later that $2 \pi E_F U$ is a half gap $m$, 
opened by the backscattering (\ref{1}) in the charge mode spectrum inside the wire.

\noindent
1.{\it High temperatures} $T>T_L,m$ - 
The average backscattering current 
$<\hat{I}_{bsc}>=\int D\phi \hat{I}_{bsc} exp\{i\int dt \int dx 
({\cal L}_c+{\cal L}_{bsc})\}$ can be written as a formal infinite series
in $U$. Each term of it is an integral of product of the free bosonic 
correlators $<exp\{ i \sqrt{2}\phi(x,t)\}exp\{ -i \sqrt{2}\phi(y,0)\}>$. 
Such a correlator approaches its uniform TLL expression when 
$x,L-x,y,L-y \gg v_c/T$. Substitution of this form into above series 
allowed us \cite{us} to find $L$-proportional part of the backscattering 
current neglecting the boundary contribution in the perturbative case.
However, the problem is not perturbative, in general, due to a finite
gap $2m$ creation. Therefore, application of the uniform correlator
will give us a part of the backscattering current $\propto min(L,v_c/m)$
with the relative error $O(max(T_L,m)/T)$, which is of the order of 
ratio
of the border piece $\propto v_c/T$ to the essential part of 
the "bulk" one. 
This relates to the high-temperature asymptotics of the whole current.
\begin{figure}[htb]
\begin{center}
\leavevmode
\psfig{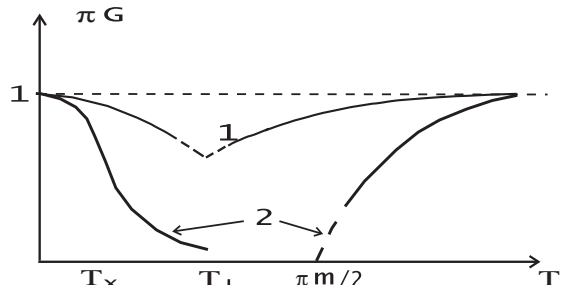}
\narrowtext{ \caption{
           Schematic linear bias conductance $G$ vs.
            temperature near half-filling $\mu \ll T_L$: 
            curve 1 for the weak interaction $m \ll T_L$,
            curve 2 for the strong one.  \label{cty}
           }}
\end{center}
\end{figure}
Calculation of above series with the uniform TLL correlator 
is equivalent to expanding the value
$g=1/2$ into the leads. 
Following Luther and Emery \cite{le}
we map this bosonic Lagrangian 
onto the free massive fermion one \cite{fuk,mas3} with the density
of Lagrangian
\endmulticols
\widetext
\vspace{-6mm}\noindent\underline{\hspace{87mm}}
\begin{equation}
{\cal L}_{F}=i\sum_{a=R,L=\pm} \psi^+_a (\partial_t \pm v_c \partial_x) \psi_a
-m \varphi(x) [e^{i(4k_Fx+2Vt)}\psi^+_L(x,t)\psi_R(x,t)+h.c.]
\label{2}
\end{equation}
\noindent\hspace{92mm}\underline{\hspace{87mm}}\vspace{-3mm}
\multicols{2}
\nopagebreak
Here $\psi_{R(L)}$ is right (left) chiral fermion field. The fermionized
backscattering current $\hat{I}_{bsc}=
2im \int_0^L dx [exp\{i(4k_Fx+2Vt)\}\psi^+_L(x,t)\psi_R(x,t)-h.c.]$ is the
doubled backscattering current for the fermions \cite{us2} under 
doubled voltage. To find its average we just need to know 
the fermionic reflection coefficient $R$
as a function of dimensionless energy $\omega = \varepsilon/m$:
\begin{equation}
R(\omega)=\frac{\sin^2(\sqrt{\omega^2-1}\bar{t}_L)}
{(\omega^2-1)+\sin^2(\sqrt{\omega^2-1}\bar{t}_L)}
\label{3}
\end{equation}
where $\bar{t}_L\equiv m/T_L$ denotes the dimensionless traversal time. 
The analytical continuation is assumed for $|\omega | < 1$.
Since the chemical potential for the right/left chiral fermions
is $\mu\pm V$, respectively,
the total current can be expressed as
\begin{eqnarray}
\lefteqn{I={V \over \pi} - {m \sinh(V/T) \over \pi} \times }
\nonumber \\
& & \hspace{5mm} \int d \omega \frac{R(\omega)}
{\cosh((m \omega-\mu)/T)+ \cosh(V/T)}
\label{4}
\end{eqnarray}
Only the leading term in $max(m,T_L)/T$ of the right hand side of (\ref{4})
is meaningful. 
Extracting it, we find the high-temperature asymptotics as following
\begin{eqnarray}
\lefteqn{
I={V \over \pi} - {m\over \pi}
\frac{ \sinh(V/T) B(mt_L)}{\cosh((\mu)/T)+ \cosh(V/T)} }
\label{5}\\
& & \hspace{3mm} B(x)\equiv \int d \omega 
\frac{\sin^2(\sqrt{\omega^2-1}x)}
{(\omega^2-1)+\sin^2(\sqrt{\omega^2-1}x)}
\nonumber
\end{eqnarray}
where function $B(x)$ increases as $\pi x$ at small $x>0$ and approaches 
the constant $\simeq \pi$ at $x\gg 1$. Accuracy of this calculation
of (\ref{4}) 
may be written as a factor $1+O(max(m,T_L)/max(T,V))$ to (\ref{5}) if 
$|\mu| \ll max(T,V)$ or as 
$1+O([max(m,T_L)/max(T,V)](max(T,V)/\mu)^2e^{|\mu|/T})$, otherwise.
The high-temperature conductance 
(Fig.\ref{cty})
\begin{equation}
G={1 \over \pi} \left( 1- {m \over T}
\frac{B(m / T_L)}
{1+\cosh(\mu/T)} \right)
\label{6}
\end{equation}
approaches zero at $T \approx m$ if the gap is large enough $m/T_L \gg 1$
and $|\mu|<m$.

\noindent
2.{\it Low energies} $T,V<<T_L$ - To find a low energy model for our 
problem we have to integrate out all high energy modes. We will try to escape
direct integration following
Wiegmann's effective way of constructing the Bethe-ansatz solvable 
models \cite{w} for the Kondo problem and for the screening of a resonant 
level  \cite{w,p}. First, let us substitute $\phi/\sqrt{2} $
instead of $\phi$ in (\ref{1}). It makes fermions interacting inside the
leads and non-interacting inside the wire. Their passage through the 
wire may be described with the one-electron $S$-matrix dependent on the
electron momentum. The interaction between electrons in the leads 
with some two-particle  $S$-matrix. Then the solution could be constructed 
if the 
proper commutation relations between the $S$-matrices are met. Being 
interested in the variation of energy less then $T_L$ around the Fermi level, 
one can simplify the solution keeping the one-particle 
$S$-matrix constant equal to its value on the Fermi level. It 
leads us to  the problem of one impurity in the TLL. 

For the weak
backscattering, the Lagrangian of this problem can be written as
\endmulticols
\widetext
\vspace{-7mm}\noindent\underline{\hspace{87mm}}
\begin{equation}
\int d x {\cal L}_t= \int dx  { v_F \over 2}
\{ 
{1 \over v_F^2} \left({{\partial_t \phi_c(t,x) }
     \over {\sqrt{4 \pi}}}
\right)^2 -
\left({{\partial_x \phi_c(t,x) } \over {\sqrt{4 \pi}}} \right)^2 
\}
- \frac{Y T_L u}{ \pi v_F} \cos(2 V t +
\sqrt{2} \phi_c(t,0))
\label{7}
\end{equation}
\noindent\hspace{92mm}\underline{\hspace{87mm}}\vspace{-3mm}
\multicols{2}
\noindent
where we  rescaled $\phi$ back and introduced a new energy cut-off parameter
$Y T_L$ with dimensionless constant $Y$ which will be specified later. 
Parameter $u$ is
related to the weak reflection coefficient as:$u^2=v_F^2 R(\mu/m)$.
For the strong backscattering the tunneling Hamiltonian approach may be applied
\cite{us3}. It was associated \cite{kf} to
the dual representation using the field $\theta$
mutually conjugated to $\phi:\  
[\theta_{\sigma}(x), \phi_{\sigma}(y)]=i 2 \pi sgn(x-y)$. 
The appropriate Lagrangian reads
\endmulticols
\vspace{-6mm}\noindent\underline{\hspace{87mm}}
\begin{equation}
\int d x {\cal L}_t= \int dx  { v_F \over 2}
\{ 
{1 \over v_F^2} \left({{\partial_t \theta_c(t,x) }
     \over {\sqrt{4 \pi}}}
\right)^2 -
\left({{\partial_x \theta_c(t,x) } \over {\sqrt{4 \pi}}} \right)^2 
\}
- \frac{Y T_L u'}{ \pi v_F} \cos(V t +
\theta_c(t,0)/ \sqrt{2}) 
\label{8}
\end{equation}
\noindent\hspace{92mm}\underline{\hspace{87mm}}\vspace{-3mm}
\multicols{2}
\noindent
with $u'^2=v_F^2 (1-R(\mu/m))$ proportional to the free massive
fermion transmittance and the voltage multiplied by $g$ factor \cite{weiss}.
Both these Lagrangian are, indeed, equivalent \cite{schm} if 
interaction dependent relation between $u$ and $u'$ is met 
\cite{fendley,weiss}. 
The above model (\ref{7}) or(\ref{8}) characterizes the  point 
scatterer of 
any backscattering strength at low energy \cite{kf}. Although, the exact relation 
between $u$ or $u'$ and the bare parameters of the scatterer
remains unknown. 
Our problem is dually symmetrical to that of Kane and Fisher:
suppression of the direct current in their problem equals suppression of 
the backscattering one in our case. This correspondence allows us to 
re-write their 
solution\cite{fendley,kf} as follows:
\begin{eqnarray}
I&=&{T_x \over \pi} Im \psi ({1 \over 2}+{T_x + i V\over \pi T})
\nonumber\\
I&=&{T_x \over \pi}\arctan(V/T_x), \ \ T=0 \label{11} \\
G&=&\frac{T_x}{\pi^2 T} \psi ' ({1 \over 2}+{T_x \over \pi T})
\nonumber
\end{eqnarray}
where $\psi$ denotes the digamma function and satisfies: $\psi'(1/2)=\pi^2/2,
\  
\psi'(x) \propto 1/x, \ x \rightarrow \infty$, and
a new energy scale $T_x$ \cite{weiss} 
varies from $T_x=Y T_L \sqrt{4R}$ at the weak backscattering (\ref{7})
to  $T_x=Y T_L (1-R)/\pi$ at the strong one (\ref{8}).

Let us, first, compare this result with the
perturbative one \cite{us,od}. The latter was derived making use of 
the long-time asymptotics for the correlator:
\endmulticols
\vspace{-7mm}\noindent\underline{\hspace{87mm}}
\begin{equation}
<e^{i \phi_c(x,t)} e^{-i \phi_c(y,0)}>=cst \bigl( {\alpha
T_L } \bigr)^{2g} \bigl({{(\pi T/T_L )^2} \over {
sinh(\pi (t-i\alpha)T) sinh(\pi (-t+i\alpha)T)}} \bigr)
F(x)F(y)
\label{9} 
\end{equation}
\noindent\hspace{92mm}\underline{\hspace{87mm}}\vspace{-3mm}
\multicols{2}\noindent
where $\alpha=1/E_F$ and $F$ was simplified: 
$F(x)=(x/L)^{gr}\prod_{\pm,m=1}^\infty(m\pm x/L)^{gr^{2m\pm 1}} \approx 
cst' \times [x(L-x)/L^2]^{gr}$.
$r=(1-g)/(1+g)$ is the reflection coefficient of the charge wave 
scattering on the contacts of the wire
with the leads. One can see that the current calculated with this
form of the correlator will be consistent with the Lagrangian (\ref{7}) 
if we choose $Y$ matching
\begin{equation}
{Y \over \sin(\mu/T_L)}=
{ cst  \over \sqrt{\pi} \Gamma(1+r)}
\frac{(2\mu/T_L)^{r-1/2}}
{J_{r+1/2}(\mu/T_L)}
\label{10}
\end{equation}
This specifies $Y$ as a constant of the order of 1 for $|\mu|/T_L < 1$. 
Then the 
solution (\ref{11}) coincides
with the perturbative result \cite{us,od}, 
that is $I-V/\pi \propto -V^3$ and $G -1/\pi \propto -T^2$.
\begin{figure}[htb]
\begin{center}
\leavevmode
\psfig{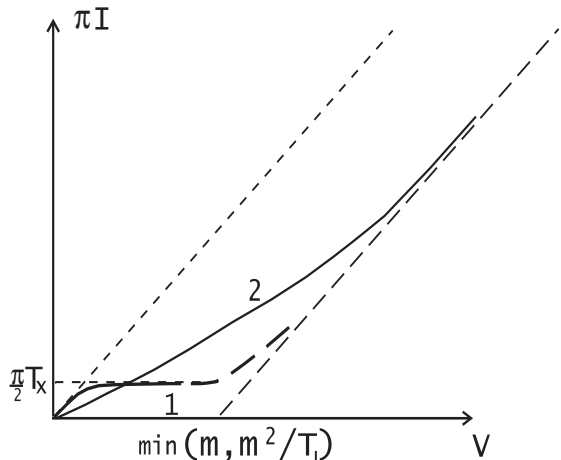}
\narrowtext{
  \caption{Schematic current $I$ vs. voltage $V$ near half-filling
           $\mu \ll T_L$: curve 1 is zero temperature 
           dependence, curve 2 is the high temperature $ T \gg T_L$ one,
           the dashed lines are the low voltage $I=V/\pi$ and 
           high voltage asymptotics.     \label{cn}
        }}
\end{center}
\end{figure}
Turning to the case $m/T_L \geq 1$ we cannot use the perturbative 
expression (\ref{10}) anymore: 
the perturbative series is not convergent due to a finite gap creation.
Then above non-perturbative consideration is necessary.
Application of the solution (\ref{11}) in this case reveals 
a quite remarkable property of low energy transport through 
the Mott-Hubbard insulator. There is
an exponentially small value of 
$T_x=(1-R(\mu/m))YT_L/\pi \propto T_L exp(-2m/T_L)$ 
for $\mu \ll m$. Hence, the zero temperature current $I$ 
(Fig.\ref{cn}) is not 
suppressed for the voltage less than $T_x$ and saturates at $T_x/2$ value 
when $T_x<V<T_L$. Similarly, 
the conductance (Fig.\ref{cty}) displays a small decrease $\propto T^2$ 
below its zero 
temperature value $1/\pi$ with increase of $T$ for $T<T_x$ 
and approaches its exponentially small asymptotics 
$G=T_x/(4T) \propto exp(-2m/T_L)T_L/T$ above $T_x<T<T_L$. As $|\mu |$ increases, 
the reflection coefficient $R(\mu/m)$ on the Fermi level
goes down and $T_x$ exceeds $T_L$, finally, approaching its weak backscattering
value $2T_x=Y T_L/\sqrt{R(\mu/m)}$, where the perturbative consideration
is applicable.

In summary, we studied transport through a 1D Mott-Hubbard insulator
beyond perturbative approach.  Assuming that $g=1/2$ near the half-filling 
in agreement with
the Bethe anztas solutions we mapped the problem onto 
the exactly solvable models and found current vs. voltage at high
temperature $T>max(m,T_L)$ and at low energy $T,V<T_L$. The solution of
these models shows, in particularly,
that the high-temperature transport through the Mott-Hubbard insulator
is similar to the one through the band gap insulator at $g=1/2$.
At low energies, however, there is always a regime where the transport
remains unsuppressed in the absence of the impurity
backscattering. 
For the strong interaction resulting in the opening
of the large Mott-Hubbard gap,
the transport through the wire is suppressed  near the half-filling almost
everywhere inside the gap except for an exponentially small 
low energy region $V,T < T_L exp(-m/T_L)$.

The authors acknowledge H. Fukuyama, S. Tarucha for useful discussions.
This work was supported by the Center of Excellence at
the Japanese Society for Promotion of Science.

$^*$ on leave of absence from A.F.Ioffe PTI, St. Petersburg, Russia

\end{document}